\documentclass[sigconf]{acmart}
\AtBeginDocument{%
  }

\usepackage{multirow}
\usepackage{subfigure}
\setcopyright{acmlicensed}
\copyrightyear{2018}
\acmYear{2018}
\acmDOI{XXXXXXX.XXXXXXX}
\acmConference[Conference acronym 'XX]{Make sure to enter the correct
  conference title from your rights confirmation email}{June 03--05,
  2018}{Woodstock, NY}
\acmISBN{978-1-4503-XXXX-X/2018/06}

\begin{document}

\title{iScript: A Domain-Adapted Large Language Model and Benchmark for Physical Design Tcl Script Generation}


% \affiliation{%

\author{Ning Xu}
\affiliation{%
 \institution{National Technology Innovation Center for EDA}
 \institution{Southeast University}
  \city{Nanjing}
  \country{China}}
\email{}

\author{Zhaoyang Zhang}
\affiliation{%
 \institution{National Technology Innovation Center for EDA}
 \institution{Southeast University}
  \city{Nanjing}
  \country{China}
}

\author{Senlin Shu}
\affiliation{%
 \institution{National Technology Innovation Center for EDA}
  \city{Nanjing}
  \country{China}
}

\author{Lei Qi}
\affiliation{%
 \institution{Southeast University}
  \city{Nanjing}
  \country{China}
}

\author{Jiaqi Lv}
\affiliation{%
 \institution{Southeast University}
  \city{Nanjing}
  \country{China}
}

\author{Wensuo Wang}
\affiliation{%
 \institution{National Technology Innovation Center for EDA}
  \city{Nanjing}
  \country{China}
}
\author{Tianhao Zhao}
\affiliation{%
 \institution{National Technology Innovation Center for EDA}
  \city{Nanjing}
  \country{China}
}

\author{Chao Zhang}
\affiliation{%
 \institution{National Technology Innovation Center for EDA}
  \city{Nanjing}
  \country{China}
}
\author{Zhaoliang Yang}
\affiliation{%
 \institution{National Technology Innovation Center for EDA}
  \city{Nanjing}
  \country{China}
}
\author{Xiangyu Li}
\affiliation{%
 \institution{National Technology Innovation Center for EDA}
  \city{Nanjing}
  \country{China}
}
\author{Zhaorui Su}
\affiliation{%
 \institution{National Technology Innovation Center for EDA}
  \city{Nanjing}
  \country{China}
}
\author{Jingshan Li}
\affiliation{%
 \institution{National Technology Innovation Center for EDA}
  \city{Nanjing}
  \country{China}
}

\author{Xin Geng}
\affiliation{%
 \institution{Southeast University}
  \city{Nanjing}
  \country{China}
}
% \affiliation{%

% \affiliation{%

% \affiliation{%

% \affiliation{%

% \affiliation{%

\renewcommand{\shortauthors}{Xu et al.}

\begin{abstract}
Modern EDA flows rely heavily on Tcl scripting, yet general LLMs perform poorly in this domain due to extreme data scarcity, domain-specific semantics, and the high reliability required in physical design. We present iScript, a domain-adapted Qwen3-8B model for Innovus Tcl script generation, and iScript-Bench, a comprehensive benchmark covering five task categories and three difficulty levels. To overcome the lack of training data, we introduce a multi-stage data synthesis pipeline that integrates command extraction, static linting, requirement back-inference, and Chain-of-Thought generation, producing a 10K-tuple (requirement, CoT, script) dataset. iScript is trained through a two-stage strategy combining domain-adaptive pretraining and supervised fine-tuning. To evaluate script correctness efficiently, we further propose a two-step verification framework consisting of static syntax verification and LLM-based functional evaluation. On our benchmark, iScript shows higher pass@k scores than currently state-of-the-art LLMs on average. These results demonstrate the effectiveness of domain adaptation and data synthesis for EDA scripting tasks.
\end{abstract}





\maketitle

\section{Introduction}
% Electronic Design Automation (EDA) tools are the cornerstone of the multi-billion dollar semiconductor industry, enabling the design of complex integrated circuits (ICs). The Place and Route (P\&R) stage, a critical part of the digital backend flow, relies on sophisticated EDA tools like Cadence Innovus to translate a synthesized netlist into a physical layout.
The physical design (PD) stage of modern ASIC implementation has grown increasingly complex as process technologies advance toward single-digit nanometers. Contemporary place-and-route (P\&R) flows rely heavily on Tcl-driven automation to integrate floorplanning, placement, clock-tree synthesis, routing, and design closure into a unified optimization pipeline. Industrial reports and tool documentation indicate that a complete production flow often requires thousands to tens of thousands of lines of Tcl scripts, covering tool configuration, constraint generation, hierarchical management, incremental optimizations, and signoff-specific tuning \cite{cadence_innovus,synopsys_icc2}. As design sizes and PPA (power–performance–area) requirements escalate, authoring and maintaining these scripts has become one of the most time-consuming and expertise-dependent tasks for PD engineers. The reliance on manually crafted Tcl scripts therefore represents not only a productivity bottleneck but also a barrier for new engineers entering advanced-node physical design.

Recent progress in LLMs has inspired interest in automating parts of semiconductor workflows. Several studies have explored LLM-assisted RTL generation \cite{liu2024rtlcoder,akyash2025rtl++}, design debugging \cite{li2025eda}, EDA document understanding \cite{qiu2024llm}, and layout analysis \cite{liu2023chipnemo,wu2024chateda,huang2025llsm}. While promising, these efforts predominantly target front-end design stages or natural-language reasoning tasks. Very few works address physical-design-specific Tcl scripting, despite its central role in industrial flows. Existing LLMs also exhibit significant limitations when applied directly to PD scripting: (1) tool-specific Tcl commands and options form a highly specialized and sparse vocabulary that general LLMs have rarely encountered; (2) PD scripts exhibit strong semantic coupling across commands (e.g., placement regions must align with floorplan geometry, CTS settings depend on timing constraints), making the mapping from natural-language requirements to executable Tcl particularly challenging; and (3) the lack of publicly available PD scripts—stemming from proprietary workflows—creates a severe data scarcity problem, hindering both model training and benchmarking. These observations highlight a fundamental mismatch between general LLMs and the domain-specific needs of physical design automation.
% A significant engineering bottleneck in this process is the reliance on Tcl (Tool Command Language) scripting. Engineers must write complex, domain-specific Tcl scripts to control every aspect of the P\&R flow, from floorplanning to clock tree synthesis and routing. This challenge is not unique to P\&R; adjacent research in analog layout highlights the "steep learning curve" and "cumbersome user experience" of proprietary scripting languages (like SKILL), creating a significant barrier to entry and reducing engineering productivity \cite{liu2025layoutcopilot}. This gap between natural language intent and formal tool-specific scripts, termed the "conversational hardware design" challenge, has been identified as a critical bottleneck in modern EDA \cite{blocklove2023chip}.
% These Tcl scripts present several challenges: \textbf{Steep Learning Curve: }The Innovus Tcl command set is vast, with thousands of commands, complex parameters, and subtle interdependencies. \textbf{Flow-Dependent Logic:} Scripts are not standalone programs; they must adhere to a strict logical design flow.
% \textbf{Opaque Documentation:} Best practices are often "tribal knowledge" learned through experience rather than clear, accessible documentation. \textbf{Time-Consuming Debugging:} A single incorrect parameter in a Tcl script can cause the tool to fail hours into a multi-day P\&R run, costing significant engineering time and computational resources.

Despite increasing interest in “EDA$+$LLM” research, the community still lacks a standardized benchmark for evaluating LLMs on PD Tcl generation. In contrast, progress in natural-language-to-code research has been driven by comprehensive benchmarks such as HumanEval \cite{chen2021evaluating}, MBPP \cite{austin2021program}, and EvalPlus \cite{liu2023your}, which define meaningful tasks, representative difficulty levels, and reproducible metrics. No similar benchmark exists for PD automation, leaving fundamental questions unanswered:

\textit{How well do state-of-the-art LLMs understand physical-design intent? Can they reliably map complex design requirements to syntactically correct and functionally meaningful Tcl commands? How do domain-adapted models compare to general models?}

Without a common benchmark, comparisons across models and reproducibility are nearly impossible.

In addition to the absence of benchmarks, there is also no consensus on how PD Tcl generation should be evaluated. True functional verification requires executing scripts inside a commercial EDA tool, which is expensive, restrictive, and non-reproducible. Conversely, purely syntax-based checks cannot assess whether a script satisfies design intent. Prior EDA$+$LLM work \cite{liu2023verilogeval} has largely sidestepped this issue, resulting in evaluations that are either incomplete (syntax-only) or qualitative (human review). A scalable, reproducible, and tool-independent evaluation framework is therefore needed to foster meaningful research progress.

To address these gaps, we present iScript, a domain-adapted LLM specialized for physical-design Tcl generation, and iScript-Bench, the first benchmark designed specifically to evaluate natural-language-to-Tcl capabilities in P\&R flows. Our work is enabled by a multi-stage data synthesis pipeline that integrates (i) domain corpus curation from publicly available sources, (ii) PD command extraction and normalization from Innovus manuals, (iii) automatic script generation using teacher LLMs with chain-of-thought reasoning, and (iv) static syntax verification via a lightweight PD sandbox. Using this pipeline, we construct a 10k-sample dataset of (requirement, reasoning, script) tuples spanning floorplanning, placement, clocking, routing, and design optimization tasks across three difficulty levels. We further apply domain-adaptive continued pre-training (CPT) and reasoning-enriched supervised fine-tuning (SFT) on top of Qwen3-8B to obtain iScript.

We also develop a two-step verification framework that combines (1) syntax checks based on a physically consistent Innovus-style sandbox, and (2) a functional evaluator LLM equipped with PD command knowledge. While not a substitute for full-tool execution, this framework enables scalable, consistent, and reproducible comparison across models.

Our contributions are as follows:
\begin{itemize}
    \item We introduce iScript, a domain-adapted LLM trained via CPT+SFT with chain-of-thought supervision, capable of generating syntactically valid and semantically meaningful PD Tcl scripts.
    \item We present iScript-Bench, the first benchmark for physical-design Tcl script generation, covering five task categories and three difficulty levels.
    \item We develop a scalable two-step verification framework that enables reliable syntax and functional evaluation without requiring access to commercial EDA tools.
\end{itemize}

\begin{figure*}[th!]
    \centering
    \includegraphics[width=1.0\linewidth]{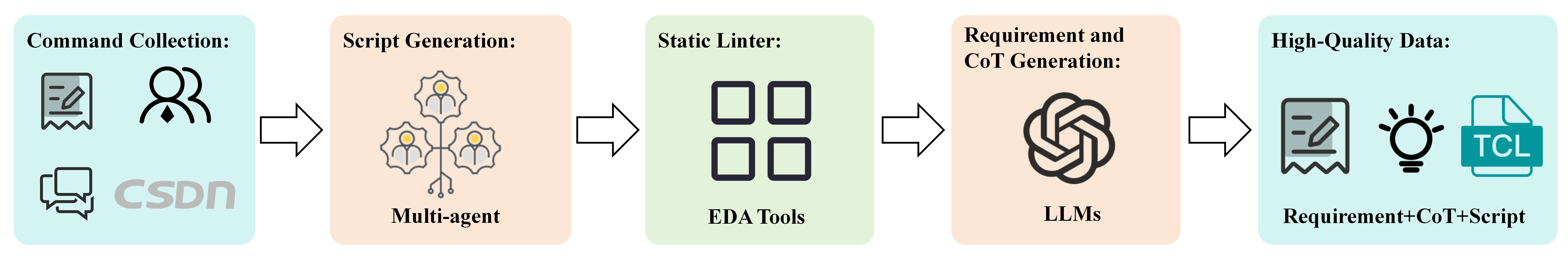}
    \caption{Training Data Synthesis}
    \label{fig:training_data}
\end{figure*}

\section{Related Works}
The application of large language models to electronic design automation spans two closely related directions: (1) general code and program synthesis, and (2) EDA-specific adaptation for design automation tasks.

\textbf{Code-oriented LLMs.} Prior work in program synthesis established the “pre-train on code, fine-tune on task” paradigm that underlies our approach. Large-scale models and datasets such as CodeTrans and CodeT5 demonstrated that syntactic and identifier-aware pretraining improves downstream code generation and retrieval performance \cite{elnaggar2021codetrans,wang2021codet5}. Benchmarks like HumanEval and MBPP provided reproducible task definitions and pass@k evaluation protocols that shaped subsequent evaluation practice in code generation communities \cite{austin2021program,chen2021evaluating}. These works motivate the (requirement $\xrightarrow{}$ script) supervised fine-tuning setup we adopt.

\textbf{Domain adaptation for chip design.} Recent efforts have shown that domain-adaptive continued pretraining on chip-design corpora yields substantial gains for EDA tasks. NVIDIA’s ChipNeMo applied continued pretraining on proprietary chip-data and demonstrated improvements in tool-script generation relative to general models \cite{liu2023chipnemo}. Similar domain-adaptation strategies have been used successfully in other verticals, motivating our CPT+SFT pipeline.

\textbf{LLMs for EDA workflows and scripting.} Several recent systems specifically target automation of EDA flows and tool control. Autonomous-agent frameworks such as ChatEDA and variations have demonstrated multi-step flow orchestration (script generation + execution) in open toolchains, highlighting both the promise and the reliability challenges of agentic automation \cite{wu2024chateda}. LayoutCopilot and related efforts have explored LLM-driven script generation for analog layout and SKILL-language tasks, emphasizing the importance of domain knowledge and interactive refinement \cite{liu2025layoutcopilot}. These works differ from ours primarily in scope: they focus on multi-agent orchestration or interactive workflows, while our work targets single-turn requirement $\xrightarrow{}$ Tcl-script generation and proposes a benchmark for systematic evaluation.

\textbf{Data synthesis and low-resource remedies.} A recurring obstacle for EDA LLMs is the scarcity of high-quality, public training data. Approaches that synthesize task data—via programmatic permutations, teacher-model back-translation, or multi-agent generation—have been used to bootstrap model training (e.g., ChipExpert \cite{chipexpert2024}, JARVIS \cite{pasandi2025jarvis}). Our data-synthesis pipeline follows this trend but emphasizes (a) command-level normalization from tool manuals, (b) static linting to ensure syntactic validity, and (c) chain-of-thought (CoT) supervision to teach reasoning about P\&R intent — a combination not previously published for Innovus-style Tcl generation.

\textbf{Evaluation methodologies.} Finally, benchmark construction and verification remain open problems. VerilogEval introduced sandbox-based functional evaluation for HDL generation, favoring execution-based correctness over lexical metrics \cite{liu2023verilogeval}. In the P\&R context, running full toolflows is costly and non-reproducible; prior works therefore resort to syntax-only checks or limited manual evaluation. Our contribution is a two-step verification strategy—static syntax checking followed by an LLM-augmented functional evaluator—designed to be scalable and reproducible while acknowledging the limitations of non-execution-based assessment.

Compared with prior works, iScript combines domain-adaptive pretraining, a synthesis-focused data pipeline with CoT supervision, and a benchmark tailored to Innovus-style Tcl tasks. The principal novelty lies in assembling these elements into a reproducible benchmark and evaluation workflow specifically for P\&R Tcl generation, together with an empirical study that compares domain-adapted and general-purpose LLMs on this task.

\section{iScript}
In this section, we introduce the training data collection, training details of iScript, and evaluation strategy.    
\subsection{Training Data Preparation}

The main challenge for training an EDA-specific LLM is the absence of a large-scale, high-quality open-source dataset. Our strategy is to leverage multi-source public data, programmatic generation, and LLM-based refinement to create a domain-specific (Instruction, CoT, Code) dataset.

\begin{figure*}
    \centering
    \includegraphics[width=1.0\linewidth]{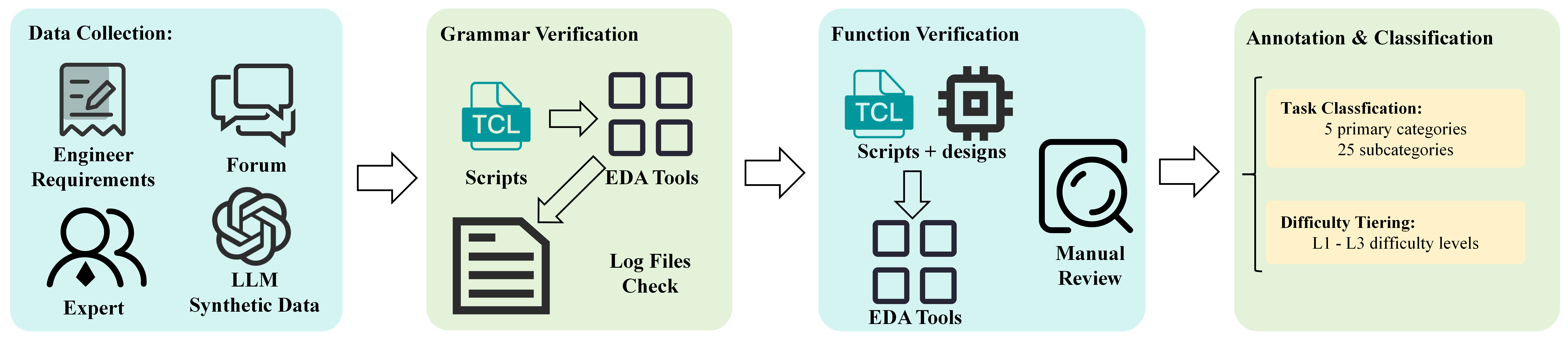}
    \caption{Benchmark Collection}
    \label{fig:collection}
\end{figure*}

As illustrated in Fig. \ref{fig:collection}, we collect seed data from a variety of sources:
\begin{itemize}
    \item EDA Tool User Guides: Provide examples of common flows and command usage.
    \item EDA Command Reference Manuals: Detail the syntax for individual commands and parameters.
    \item Chinese Technical Communities (CSDN): Offer tutorials and user-generated script examples.
    \item Technical Forums: Contain user questions and expert-provided solutions.
\end{itemize}

Our pipeline consists of three novel stages:

\textbf{Stage 1: Script Generation via Combination}
Firstly, we extract a comprehensive list of core Tcl commands and their associated parameters from the collected data. We then generate a vast corpus of Innovus Tcl script fragments by creating permutations and combinations of these commands and parameters. The generation is based on a multi-agent system inspired by \cite{pasandi2025jarvis}. This initial corpus is large but of low quality and often syntactically incorrect.

\textbf{Stage 2: Static Syntax Validation}
The raw scripts from Stage 1 are passed through a Tcl Linter (static syntax checker). This step is crucial as it filters out all syntactically invalid scripts, ensuring that the data used for the next stage is at least structurally correct. This results in a large corpus of syntactically valid scripts.

\textbf{Stage 3: Requirement Back-Inference and CoT Data Generation}
This is the core of our data synthesis contribution.
\begin{itemize}
    \item Requirement Back-Inference: We use a powerful teacher model (GPT-4.1) to "reverse-engineer" the syntactically valid scripts. The model is prompted to infer the most likely natural language "requirement" or "intent" that would produce the given script.
    \item CoT generation: We prompt the teacher model a second time using the (Requirement, Script) pairs. We ask it to generate a Chain-of-Thought (CoT) explanation, detailing the step-by-step reasoning for how the script fulfills. 
\end{itemize}
% In \textit{requirement back-inference}: We use a powerful teacher model (GPT-4.1) to "reverse-engineer" the syntactically valid scripts. The model is prompted to infer the most likely natural language "requirement" or "intent" that would produce the given script. In \textit{CoT generation},  we prompt the teacher model a second time using the (Requirement, Script) pairs. We ask it to generate a Chain-of-Thought (CoT) explanation, detailing the step-by-step reasoning for how the script fulfills. 

This three-stage process transforms low-quality, multi-source data into a high-quality (Requirement, CoT, Script) dataset suitable for SFT, which we call our "High-Quality Q\&A" data. This pipeline produced our final dataset, consisting of 10,000 high-quality (Requirement, CoT, Script) tuples. The dataset primarily targets Cadence Innovus Tcl and covers tasks including floorplanning, power planning, placement, clock tree synthesis, and routing.

\begin{figure*}[th]
    \centering
    \includegraphics[width=1.0\linewidth]{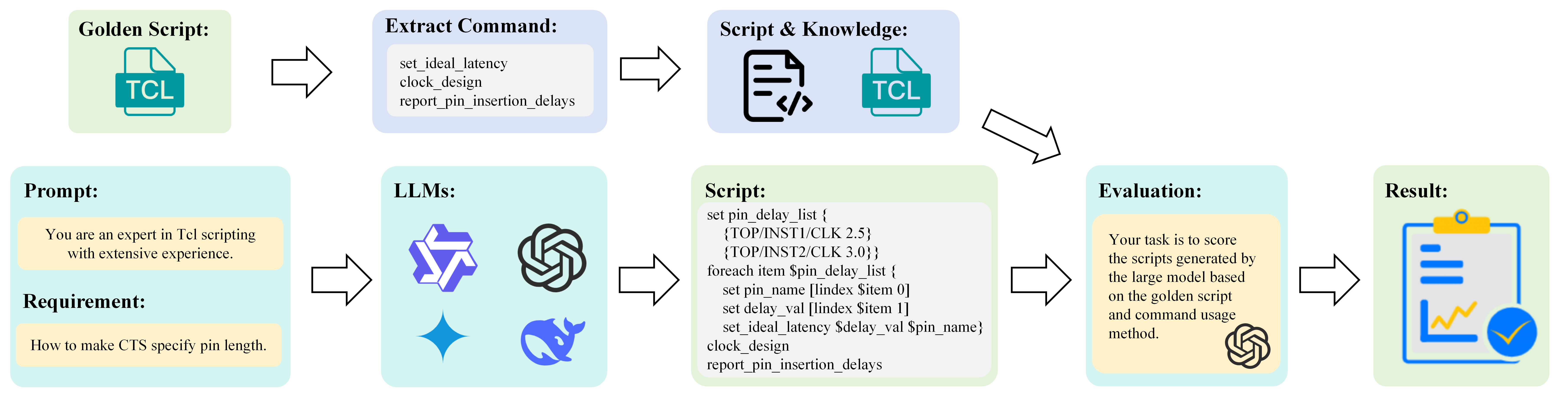}
    \caption{Evaluation Workflow.}
    \label{fig:evaluation}
\end{figure*}

\subsection{Training Details}
We selected Qwen3-8B \cite{yang2025qwen3} as our base model. This decision was based on its strong performance in general coding benchmarks, its native bilingual (English/Chinese) capabilities which align well with our multi-source data (CSDN, forums), and its 8B parameter size, which offers a robust balance between high performance and feasible training costs.
Our training follows the state-of-the-art "CPT$+$SFT" methodology, which has been proven effective for domain-specific EDA tasks by \cite{liu2023chipnemo} and ChatEDA \cite{wu2024chateda}.

\textbf{Domain-Adaptive Continued Pre-Training (CPT)} The primary goal of this stage is to embed the comprehensive "vocabulary" and syntactical patterns of Innovus Tcl into the model's weights. The base Qwen3-8B model has likely seen very little, if any, Innovus Tcl data in its general pre-training. This CPT stage "warms up" the model, adapting its internal representations to this new, specialized language. When the model enters Stage 2, it will no longer struggle with how to write Tcl; it only needs to learn which Tcl to write in response to an instruction. After CPT, the model gains high-level proficiency in the Innovus Tcl syntax. Its ability to generate coherent, syntactically valid script snippets (its "fluency") is significantly improved, providing a robust foundation for the next stage.

\textbf{Supervised Fine-Tuning (SFT) with Chain-of-Thought} With the model now fluent in Tcl syntax, this stage teaches it the high-level semantic mapping from a user's intent (the Requirement) to the correct action (the Script). We explicitly include Chain-of-Thought (CoT) data in this stage to teach the model not just what to generate, but why. Research such as LLSM \cite{huang2025llsm} has shown that "EDA-guided CoT" is highly effective for complex reasoning in the EDA domain (specifically for logic synthesis). Furthermore, EDAid \cite{wu2025divergent} everages CoT to improve the reliability and reasoning of their multi-agent system. By training our model to generate this reasoning path, we improve its ability to handle complex, multi-step requests and adhere to logical P\&R flows. The final EDA-Gen model is capable of both understanding natural language requirements for P\&R and generating syntactically correct, functionally appropriate Innovus Tcl scripts, complete with a step-by-step reasoning trace. This two-stage approach is expected to significantly outperform baseline models that have not undergone this specialized domain adaptation.

\textbf{Implement Details} 
Training was performed using the LLaMA Factory \cite{zheng2024llamafactory} framework on a cluster of eight NVIDIA A800 GPUs. The training procedure consisted of two stages. In the initial Domain-Adaptive Continued Pre-Training (CPT) phase, the model was trained for 2 epochs on a corpus of raw Tcl scripts with a learning rate of $2 \times 10^{-5}$, a global batch size of 64, and a weight decay of 0.1. Following this, in the Supervised Fine-Tuning (SFT) phase, the model was trained for 1 epoch with a global batch size of 64, a learning rate of $5 \times 10^{-6}$, and a weight decay of 0.1.

\subsection{Evaluation Strategy}

A comprehensive benchmark and automated evaluation system are important components for measuring the code generation task of large models. However, in EDA script generation, such benchmark and system are missing. To address this issue, on one hand, we have proposed a benchmark that covers a wide range of fields and has multiple difficulty levels. On the other hand, due to the long verification time of a complete design and the inability of a design to meet all requirements, it is difficult to establish a system that can fully automatically verify script functionality. Therefore, we propose a two-step verification strategy, starting with static syntax detection and then conducting functional evaluation based on LLM.

\subsubsection{iScript-Bench}

In order to establish a real and comprehensive benchmark, we collected and organized scripts from various sources including expert writing, real-life needs, and the community. After the script collection is completed, we will perform static syntax checks and manual functional checks on these scripts. The processing flow of the benchmark is shown in Figure 3. Meanwhile, based on the different types of requirements and the actual difficulty of the scripts, we have divided the scripts that have passed the manual detection function into 5 primary categories, 25 subcategories, and 3 difficulty levels. The difficulty classification information is shown in Table \ref{tab:tiering}. As demonstrated in Table \ref{tab:tiering}, the benchmark features a balanced difficulty distribution. This design facilitates a dual-level assessment: it not only measures a model's comprehension and application of fundamental commands and parameters but also evaluates its proficiency in handling complex scripting logic and intricate command sequences. The classification information is as follows:
\begin{itemize}
    \item \textbf{Design Information Query and Adjustment (DIQA)}: querying various types of objects in the design, including clock paths, data paths, and objects with specific features. 
    \item \textbf{Netlist Information Analysis and Adjustment (NIAA)}: retrieving net properties, checking net connectivity, adjusting net parameters.
    \item \textbf{Floorplan Adjustment (FA)}: analysis and optimization of macros, global layout, and ports.
    \item \textbf{Placement Analysis and Optimization (PAO)}: analysis and constraint optimization of standard cell placement.
    \item \textbf{Timing Analysis and Optimization (TAO)}: Adjusting the topology and parameters on the clock and data paths for optimization.
\end{itemize}
These five task categories encompass the principal operations within the floorplanning stage, enabling a comprehensive evaluation of the model's ability to comprehend these operations and generate syntactically correct and functionally appropriate code.

\begin{table*}[th!]
    \centering
    \caption{Difficulty Tiering}
    \label{tab:tiering}
    \begin{tabular}{ccc}
    \toprule
         Level &  Number & Description\\
         \hline
            L1  &  54      &    single line of code or the simple use of a single command. \\
            L2  &  36      &    combination of multiple commands and their parameters \\
            L3  &  26     &    not only the combination of multiple commands and parameters but also require foundational code logic\\
    \bottomrule
    \end{tabular}
\end{table*}

\begin{figure}
    \centering
    \subfigure[]{\includegraphics[width=0.48\linewidth]{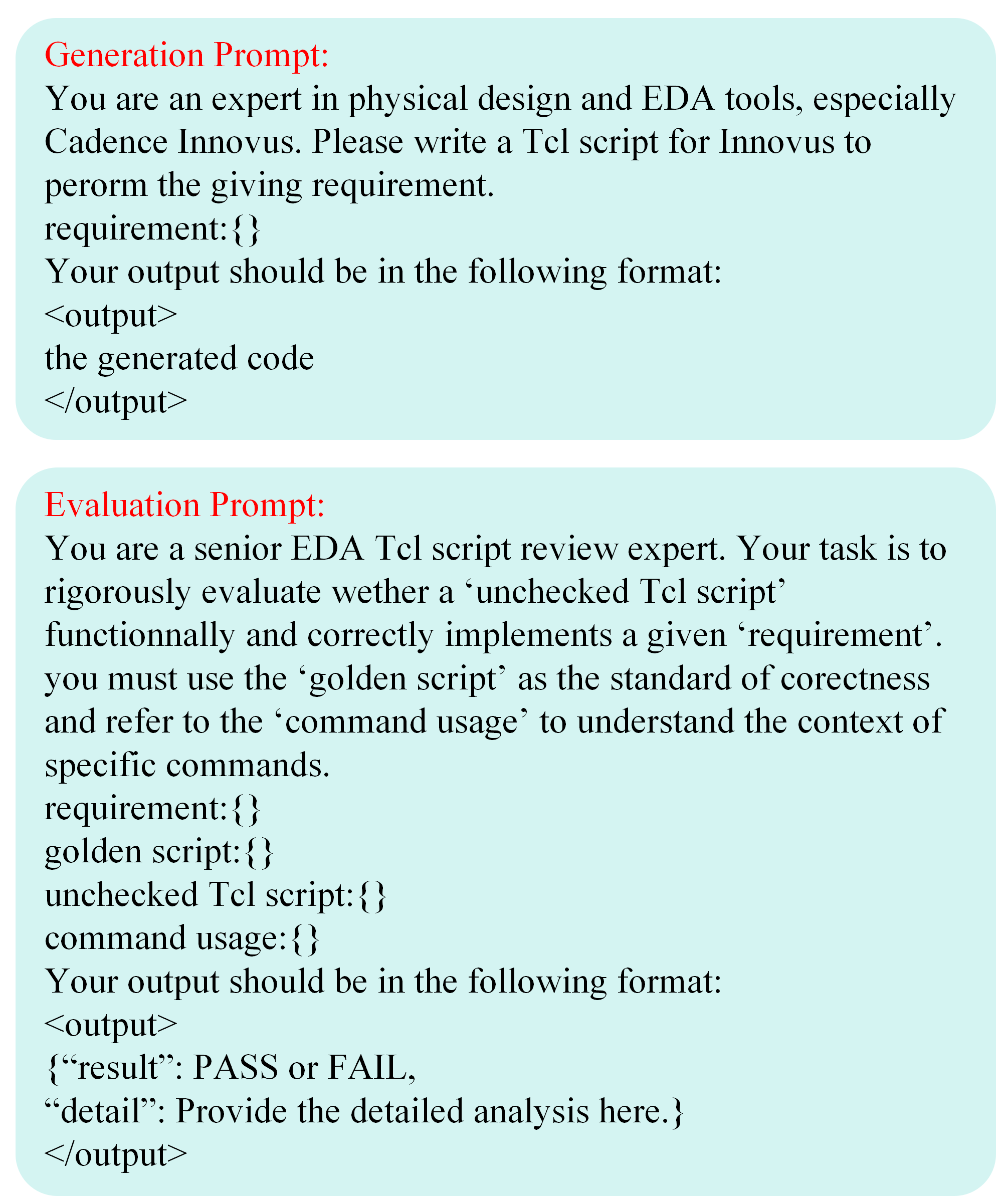}}    
    \subfigure[]{\includegraphics[width=0.48\linewidth]{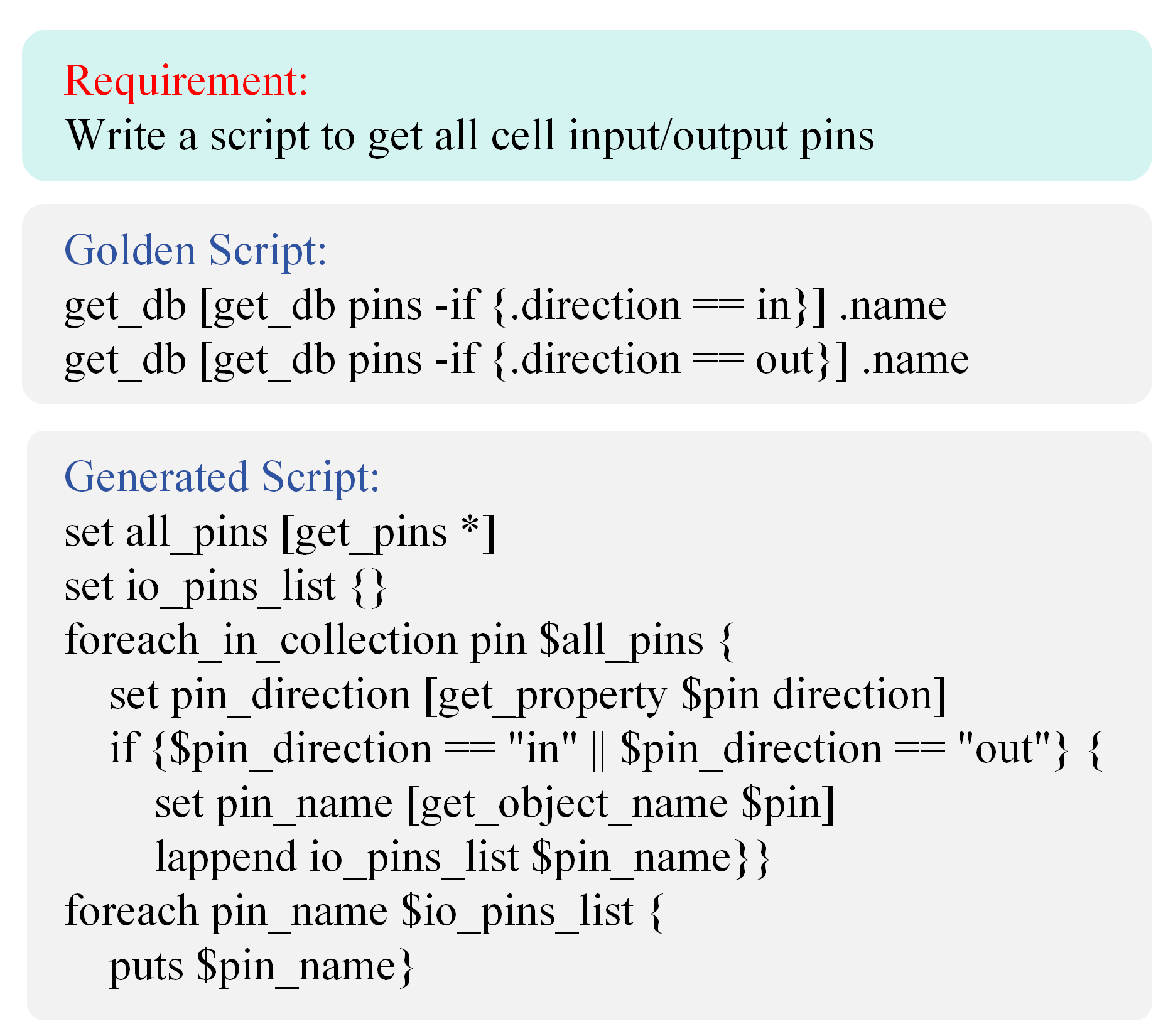}}     
    \caption{(a) Prompt Template and (b) Evaluation Example.}
    \label{fig:prompt_and_example}
\end{figure}
\subsubsection{Two-Step Verification Strategy}

% \begin{figure}
%     \centering
%     \includegraphics[width=0.5\linewidth]{example.png}
%     \caption{Evaluation Example.}
%     \label{fig:eval_example}
% \end{figure}
The evaluation of the accuracy of the EDA scripts presents a unique challenge. A fully automated, execution-based frame-work is often intractable, while purely manual grading is unscalable. Therefore, we propose a two-step verification strategy, including static syntax verification and LLM-based Functional evaluation. 

\textbf{Static Syntax Verification}: While establishing a fully automated functional verification system for EDA scripts is computationally prohibitive, ensuring syntactic correctness is a feasible and critical first step. We designed a lightweight sandbox environment based on a minimal physical design layout to perform static syntax checks. In this process, generated scripts are executed within the sandbox, and the resulting tool logs are parsed. Specifically, if the log contains the pattern "** ERROR:", the script is flagged as syntactically invalid; otherwise, it is deemed syntactically correct.

\textbf{LLM-based Functional Evaluation}: Scripts that pass the static syntax check proceed to an LLM-based functional evaluation. A primary challenge in this stage is ensuring the reliability of the LLM's judgment. To address this, we employ a knowledge-enhanced prompting strategy. We provide the evaluator LLM with not only the original design requirement and the "golden" (reference) script but also precise command definitions and object property descriptions extracted from official user manuals. This augmented context enables the model to accurately comprehend the programming logic and verify if the generated script fulfills the intended function. To further validate the reliability of this metric, we analyze the alignment between LLM judgments and human expert reviews, as detailed in Section 4.

\begin{table*}[]
    \centering
    \caption{Pass@1 (\%) on different Categories.}
    \label{tab:class_pass1}
    \begin{tabular}{c|cc|cc|cc|cc|cc|cc}
    \toprule
         Tasks &  \multicolumn{2}{c|}{DIQA (35)}  &  \multicolumn{2}{c|}{NIAA (10)}&  \multicolumn{2}{c|}{FA (22)} & \multicolumn{2}{c|}{PAO (21)} & \multicolumn{2}{c}{TAO (28)}  &\multicolumn{2}{c}{Total (116)}\\
         \hline
  &syntax   &function & syntax   &function   & syntax  &function  & syntax   &function  & syntax  &function  & syntax  &function\\
          \hline
          GPT    &8.57    & 8.57         &0.00      &0.00      &0.00      &0.00      &4.76      & 0.00       &17.86     &0.00  &7.76          &2.59 \\
          Gemini &31.43  &17.14         &\textbf{60.00}  &\textbf{30.00} &22.73  &18.18 &28.57  &9.52 &28.57  &7.14 &31.03          &14.66 \\
          Claude    &31.43 &17.14 &50.00  &\textbf{30.00} &0.00  &0.00 &9.52  &4.76 &14.29  &7.14 &18.97          &10.34\\
          DeepSeek  &11.43 &2.86 &50.00  &20.00 &4.55  &0.00 &9.52  &0.00 &3.57  &0.00 & 11.21         &2.59 \\
          iScript   &\textbf{71.43} &\textbf{25.71} &30.00  &10.00 &\textbf{77.27}  &\textbf{22.73} &\textbf{66.67}  &\textbf{9.52} &\textbf{35.71}  & \textbf{17.86} &\textbf{59.48}          &\textbf{18.97} \\
    \bottomrule
    \end{tabular}
\end{table*}

\begin{table*}[]
    \centering
    \caption{Pass@5 (\%) on different Categories.}
    \label{tab:class_pass5}
    \begin{tabular}{c|cc|cc|cc|cc|cc|cc}
    \toprule
         Tasks &  \multicolumn{2}{c|}{DIQA (35)}  &  \multicolumn{2}{c|}{NIAA (10)}&  \multicolumn{2}{c|}{FA (22)} & \multicolumn{2}{c|}{PAO (21)} & \multicolumn{2}{c}{TAO (28)}  &\multicolumn{2}{c}{Total (116)}\\
         \hline
           &syntax &function & syntax &function & syntax &function & syntax &function & syntax &function & syntax &function\\
          \hline
          GPT &45.71 &25.71 &30.00  &0.00 &0.00  &0.00 &19.05  &9.52 &28.57  &0.00 &26.72  &9.48 \\
          Gemini &85.71 &57.14 &\textbf{80.00}  &40.00 &40.91  &27.27 &66.67  &28.57 &\textbf{85.71}  &35.71 &73.28  &39.66 \\
          Claude &60.00 &45.71 &60.00  &\textbf{50.00} &9.09  &0.00 &28.57  &14.29 &21.43  &10.71 &35.34  & 23.28\\
          DeepSeek &40.00 &17.14 &60.00  &30.00 &4.55  &0.00 &23.81  &4.76 &28.57  &0.00 &29.31  &8.62 \\
          iScript &\textbf{100.00} &\textbf{71.43} &\textbf{80.00}  &20.00 &\textbf{95.45}  &\textbf{40.91} &\textbf{100.00}  &\textbf{33.33} &75.00  &\textbf{39.29} &\textbf{91.38}  &\textbf{46.55} \\
    \bottomrule
    \end{tabular}
\end{table*}

\begin{table*}[]
    \centering
    \caption{Pass Rate (\%) on different difficulties.}
    \label{tab:level_pass5}
    \begin{tabular}{c|cccc|cccc|cccc}
    \toprule
         Difficulty &  \multicolumn{4}{c|}{L1}  &  \multicolumn{4}{c|}{L2}&  \multicolumn{4}{c}{L3} \\
         \hline
           &\multicolumn{2}{c}{syntax} &\multicolumn{2}{c|}{function} &\multicolumn{2}{c}{syntax} &\multicolumn{2}{c|}{function} &\multicolumn{2}{c}{syntax} &\multicolumn{2}{c}{function}  \\
          Metric      &Pass@1  &Pass@5  &Pass@1  &Pass@5   &Pass@1   &Pass@5  &Pass@1   &Pass@5   &Pass@1    &Pass@5   &Pass@1   &Pass@5  \\
          \hline
          GPT      &9.26    &29.63   &5.56    &16.67    &11.11    &36.11   &0.00     &5.56     &0.00      &15.38    &0.00     &0.00\\
          Gemini  &25.93   &74.07   &12.96   &59.26    &41.67    &72.22   &\textbf{22.22} &27.78 &26.92  &73.08    &7.69     &15.38\\
          Claude      &27.78   &37.04   &20.37   &33.33    &16.67    &44.44   &2.78     &22.22    &3.85      &19.23    &0.00     &3.85\\
          DeepSeek    &12.96   &38.89   &3.70    &12.96    &13.89    &27.78   &2.78     &8.33     &3.85      &11.54    &0.00     &0.00\\
          iScript     &\textbf{66.67} &\textbf{94.44} &\textbf{27.78}  &\textbf{68.52} &\textbf{47.22}  &\textbf{88.89}   &11.11 &\textbf{33.33} &\textbf{61.54}  &\textbf{88.46} &\textbf{11.54}  &\textbf{19.23}\\
    \bottomrule
    \end{tabular}
\end{table*}

\section{Experiments}
\subsection{Experimental Setup}
\noindent\textbf{Model Selection}: In the EDA script generation, we compare our proposed iScript with four state-of-the-art LLMs, including GPT-4.1 \cite{achiam2023gpt}, Gemini-2.5-pro \cite{comanici2025gemini}, Claude-sonnet-4.5, and DeepseekV3.1 \cite{liu2024deepseek}. In LLM-based functional evaluation, we choose Gemini-2.5-pro with strong comprehensive capabilities as the judgment model. For those models, we employ the temperature to 1 in the script generation and use the latest version of API. 

\noindent\textbf{Prompt Template}: For the prompt words, we have defined roles, tasks, and specified the output format. Please refer to Figure \ref{fig:prompt_and_example} (a) for detailed information.

\noindent\textbf{Pass Rate}: We employ the commonly used pass@k as the evaluation indicator. We employ pass@1 and pass@5 to evaluate the performance of these LLMs and our iScript.

\subsection{Experimental Results}
\noindent\textbf{Results of Different Categories}: In order to evaluate the ability of various models in different types of design tasks, we conducted experiments on different categories of tasks. The results are shown in Table \ref{tab:class_pass1} and Table \ref{tab:class_pass5}. As shown in Table \ref{tab:class_pass1} and Table \ref{tab:class_pass5}, iScript consistently outperforms all other general LLMs (GPT, Gemini, Claude, and DeepSeek) across nearly all categories. In terms of the Pass@1 metric (Table 2), iScript achieves the highest Total scores with 59.48\% in syntax and 18.97\% in functional correctness. This is a significant lead over the second-best performer, Gemini, which scores 31.03\% and 14.66\% respectively. iScript demonstrates particular strength in the FA and DIQA categories, where other models like GPT and DeepSeek often struggle to achieve any functional correctness (0.00\%). Looking at the Pass@5 metric (Table 3), the performance gap widens further. iScript achieves a remarkable Total syntax score of 91.38\% and a functional score of 46.55\%. 

\noindent\textbf{Results of Different Levels}: In order to evaluate the ability of various models on design tasks of different difficulty levels, we conducted experiments on tasks of different difficulty levels. The results are shown in Table \ref{tab:level_pass5}. As shown in Table \ref{tab:level_pass5}, performance generally declines as the difficulty increases from L1 to L3, yet iScript maintains superior robustness compared to other models. At the L1 (easiest) level, iScript dominates with a Pass@1 syntax rate of 66.67\% and a Pass@5 rate of 94.44\%. As the difficulty rises to L3 (hardest), most general models see a drastic drop in performance; for instance, Claude and DeepSeek drop to roughly 3.85\% in Pass@1 syntax. In contrast, iScript maintains a high Pass@1 syntax score of 61.54\% at level L3. Although functional correctness remains a challenge for all models at the L3 level (with most scoring 0.00\% in Pass@1 functional accuracy), iScript still manages to secure the lead with 11.54\%. This indicates that while complex logic remains difficult for all models, iScript is significantly better optimized for handling high-difficulty design tasks than its competitors.

\noindent\textbf{Competition between Human and LLM}: To validate the reliability of our LLM-based functional evaluation, we conducted a comparative study with manual verification. We randomly sampled 100 generated scripts from the pool of 784 syntactically valid outputs and submitted them to experienced engineers for review. The results show that the LLM and human experts identified 39 and 42 functionally correct scripts, respectively. Crucially, the set of 39 scripts approved by the LLM is a complete subset of the 42 human-verified scripts. This inclusion relationship demonstrates that our automated evaluator achieves near-perfect precision (no false positives), serving as a rigorous and trustworthy lower-bound estimator of model performance. While the LLM evaluator is conservative, its high reliability ensures that the reported pass rates are not inflated. Upon analyzing the discrepancies (i.e., the 3 scripts rejected by the LLM but approved by experts), we found that these scripts achieved the required functionality using alternative commands distinct from the golden reference. Figure \ref{fig:prompt_and_example} (b) illustrates a representative example of such a false negative, where the model strictly adhered to the reference logic and penalized a valid variation.

\section{Conclusion}
For EDA script generation, there are three main challenges, including extreme data scarcity, lack of proprietary models, and lack of efficient evaluation methods. To address those challenges, we proposed iScript, a specific LLM for EDA script generation via two-stage-training. Specifically, we first employ a data synthesis framework based on multi-agent to generate massive high-quality (requirement+CoT+script) pair. Then, we have trained iScript via a two-stage training process on the based model Qwen3-8B. Moreover, to effectively and correctly evaluate the generation performance, we proposed a two-step verification strategy, static syntax verification and LLM-based functional evaluation. Existing experimental results shown the effectiveness of our iScript. 

\noindent\textbf{Limitation}: Overall, our work have two main limitations. Firstly, the existing amount of training data is still slightly insufficient, resulting in insufficient understanding of the combination between command codes for iScript to handle complex tasks. Secondly, the LLM-based functional evaluation has shown effectiveness compared to human evaluation, but an auto evaluation system is the optimal choice, since it would demonstrate the actually executed results of the script. Therefore, our future work is collect more high-quality data and establish a auto evaluation system.

% \begin{acks}
% \end{acks}

\newpage
\bibliographystyle{ACM-Reference-Format}
\bibliography{sample-base}

\end{document}